\documentclass[aps,twocolumn,notitlepage,superscriptaddress]{revtex4-2}

\pdfoutput=1

\usepackage{graphicx,amsmath,amssymb,braket,bbm,bm,mathtools}
\usepackage{latexsym}
\usepackage{color,braket}
\usepackage{titlesec}
\usepackage{dsfont}
\usepackage[tight]{subfigure} 
\usepackage[papersize={8.5in,11in}]{geometry}
\usepackage{color}
\definecolor{darkblue}{rgb}{0.,0.,0.4}
\definecolor{darkred}{rgb}{0.5,0.,0.}
\definecolor{BlueViolet}{RGB}{138,43,226}
\definecolor{SkyBlue}{RGB}{30,144,255}
\definecolor{DarkGreen}{RGB}{0,100,0}
\usepackage[pdftex,colorlinks=true,linkcolor=darkblue,citecolor=blue,urlcolor=darkred]{hyperref}
\def \nn{\nonumber \\}
\newcommand{\tr}{\operatorname{tr}}

\begin{document}
	
\title{Circular dichroism as a probe for topology in three-dimensional semimetals}

\author{Sajid Sekh}
\affiliation{Institute of Nuclear Physics, Polish Academy of Sciences, 31-342 Krak\'{o}w, Poland}

\author{Ipsita Mandal}
\affiliation{Institute of Nuclear Physics, Polish Academy of Sciences, 31-342 Krak\'{o}w, Poland}
\affiliation{Department of Physics, Stockholm University, AlbaNova University Center,
106 91 Stockholm, Sweden}

\begin{abstract}
Higher-pseudospin fermions, associated with multiple band-crossings in topological semimetals, are the condensed matter analogues of higher-spin fermions in high-energy physics. In this paper, we demonstrate that analyzing the response of a circular drive is an effective way to detect the topology of the lowest-energy Bloch band, as it can be connected to a frequency-dependent probe. The dichroic response exhibits circular dichroism due to the differential excitation rates by the left- and right-circular orientations of a time-periodic drive, induced on a filled band, because of the geometrical properties of the Bloch bands. Our analytical approximation reveals that the dichroic response is quantized for isotropic systems, when the frequency of the drive is above a critical value, and thus correctly infers the ground-state Chern number. We demonstrate this through explicit numerical computations by considering three kinds of semimetals with pseudospin values of 1/2, 1, and 3/2, respectively, and all having linear dispersions. Furthermore, we investigate the effects of tilt and anisotropy on the systems, and find that although tilt does not have any effect on the response, the presence of anisotropy can drastically hamper the quantization. Our scheme thus provides an important methodology for designing future experiments to detect the topology of band structures.
\end{abstract}

\maketitle


\section{Introduction}
Recently, there has been a surge of interest in gapless topological phases that arise in multi-band fermionic systems harbouring a band-crossing point \cite{bradlyn,bernevig2} in the Brillouin zone (BZ), where two or more bands cross and have nonzero Chern numbers. Some of these have high-energy counterparts (e.g. Dirac and Weyl semimetals). Since the prediction of these topological semimetals, a plethora of efforts has been made through the last decade to detect and characterize their topological signatures in real materials. This requires measurement of topological invariants, which often necessitates novel experimental strategies, such as quantum interference~\cite{dai}, resonant x-ray probing~\cite{stefanos} or neutron scattering~\cite{turner}. For systems like Weyl semimetals (WSMs), one of the promising ideas has been the utilization of its chiral anomaly.
This can be measured as the chiral magnetic effect (CME), where the hallmark signature involves a current, as the DC limit of the response to a time-dependent magnetic field \cite{chen,goswami}. 
Other proposals~\cite{juan,juan2,chan,kozii,Mandal_2020} include measurements of non-linear responses like circular photogalvanic effect (CPGE), or studying the trace of quantum metric from the spread of Wannier functions~\cite{lin}. 

In this paper, we focus on a scheme based on circular dichroism \cite{souza,valley-dep,PhysRevB.93.205133,PhysRevB.97.035153,goldman_dir,kush}, which probes the topology of the bands by measuring the excitation rate of a quantum state. Initially proposed for two-dimensional (2D) systems, this method involves subjecting the lowest Bloch band (LBB) to a circular time-periodic perturbation (or circularly-polarized light), with frequency $\omega$, which generates a chirally sensitive response $\gamma^{\pm}(\omega)$. Here, $\gamma^{\pm} (\omega )$ (with $\pm$ referring to the orientation or chirality of the drive) denotes the extraction rate of particles from the populated band, and can be measured experimentally by counting the total number of particles [$N_{\pm}(t)\sim \gamma^{\pm}(\omega) \,t$] scattered out of the LBB, after an elapsed time $t$. Whenever the LBB is characterized by a non-trivial Chern number $\mathcal{C} _0$, we have $N_+ \neq N_-$, that gives a nonzero net contribution $\Delta \gamma = (\gamma^+-\gamma^-)/2$.
For a 2D system, integrating $\Delta \gamma$ over the BZ, as well as a relevant frequency range, we get a quantized response (proportional to $\mathcal{C} _0$), dubbed as the differential integrated rate in Ref.~\cite{goldman_dir}.
This result crucially hinges on the differential excitation rates induced on the LBB, by the left- and right-circular orientations of the drive, because of the geometrical properties of the Bloch bands. 

Following the above scheme, we build a proposal to decipher band topology in three-dimensional (3D) momentum space. For a given plane of polarization, we compute $\Delta \gamma$ for a transition from the LBB to a higher-energy band, and multiply it with the corresponding difference in band velocities. Summing this quantity over all such allowed transitions gives a differential current, directed perpendicular to the plane of polarization. Finally, summing the contributions from differential currents in the three mutually perpendicular directions, and integrating over the 3D BZ, we get the 3D differential integrated rate per unit area (which we abbreviate as DIR), denoted by $\Gamma^{int} (\omega)$,
which encodes the topology of the Bloch bands. For a system with isotropic energy dispersions, $\Gamma^{int} (\omega)$ turns out to be quantized above a critical frequency value $\omega_c$.
 
Plane-polarized circular drives can be realized through circular shaking in ultracold atoms trapped in optical lattices~\cite{eckardt}, using piezo-electric actuators~\cite{jotzu}. Recently, this method has been implemented to measure the Chern numbers of the two-band Haldane model~\cite{goldman_dir}, and topological floquet bands~\cite{asteria}. Furthermore, a similar technique with linear shaking has been proposed~\cite{ozawa} to probe the quantum metric tensor in periodically driven systems. 

In particle physics, the Rarita-Schwinger (RS) equation represents the relativistic field equation of spin-3/2 fermions. Unfortunately, RS particles do not appear in the standard model, and have not been observed in any high-energy experiments or cosmological observations. Unlike the Poincar\'e symmetry in high-energy physics, the quasiparticles in condensed matter physics are constrained by the rich space-group symmetries, providing a pathway for potentially realizing exotic excitations that do not have a high-energy counterpart. Examples include multifold semimetals, which harbour excitations of pseudospin-1~\cite{bradlyn} and pseudospin-3/2~\cite{long,igor2,ips3by2,ma,tang}. Pseudospin-3/2 excitations are often called the Rarita-Schwinger-Weyl (RSW) fermions, as they are the condensed matter analogues of the elusive high-energy RS fermions. 
The discovery of such multifold semimetals is of great importance, as it enables researchers to probe excitations with different pseudospin (analogue of the real spin) values, in the form of massless quasiparticles within the low-energy limit.
For instance, Dirac and Weyl quasiparticles have already been observed in graphene~\cite{novoselov} and TaAs semimetals~\cite{hasan}, respectively, and there exist clear-cut theoretical predictions for the emergent Majorana quasiparticles at the edges of a topological nanowire~ \cite{kitaev,ips-sudip,ips-tewari,ips_epl_majorana,ips_counting_majorana,ips-jay}. Furthermore, the signatures of the RSW fermions have recently been associated with the large topological charges found in various materials like CoSi~\cite{takane}, RhSi~\cite{sanchez}, AlPt~\cite{schroter}, and PdBiSe~\cite{ding}.

The central idea behind realizing such massless quasiparticles is rooted in the low-energy expansion of the Hamiltonian around the nodal points, which can be described as
\begin{align} 
	\label{eq:kdots}                                  
	\mathcal{H}= \hbar \, v_F \, \delta \mathbf{k}\cdot \mathbf{S} \,, 
\end{align}
where $\hbar$ is the reduced Planck's constant, $v_F$ is the Fermi velocity,
and $\mathbf{\delta k=k-k_0 }$ denotes the momentum with respect to a band-touching point at $\mathbf k =\mathbf{k_0}$. Furthermore, $\mathbf{S}$ is a three-vector, made up of the matrices $S_j$ obeying the algebra $\left [S_j,S_k \right ]=i\,\epsilon_{jkl} \,S_l$ [with $j, k, l \in (1,2,3) $], that represents the pseudospin degrees of freedom.
For example, the Hamiltonian for the pseudospin-1/2 Weyl quasiparticles (with two nodes and opposite topological charges $\pm 1$) is governed by the $2\times2$ Pauli matrices $\boldsymbol{\sigma}$, such that $\mathbf{S}= \frac{1}{2}\boldsymbol{\sigma}$. The topological charge at a nodal point of a semimetal simply indicates the total Chern number of either the positive energy or negative energy bands (the zero of the energy being taken at the nodal / band-touching point). 
This topological charge is the manifestation of band-touching at nodal points, which acts as a source or sink of Berry flux. As a result, higher-pseudospin systems provide a fertile playground for realizing materials with higher-order topological charges. 
A prime example is the four-band RSW model with Chern numbers $\pm 1, \pm 3$, which can thus harbour topological charges of $\pm 4$.
In this paper, we focus on the $\Gamma^{int}$ response of these 3D semimetals having pseudospin-3/2 quasiparticle excitations. We also consider WSMs and pseudospin-1 systems, exhausting the cases of 3D topological semimetals with linear dispersions. The pseudospin-1 system is sometimes referred to as Maxwell fermions \cite{PhysRevA.96.033634}, and has a flat band in addition to two linearly dispersing bands. Last but not the least, we investigate the effects of tilt and anisotropy on the $\Gamma^{int}$ features. 

The paper is organized as follows. In Sec.~\ref{secmodel}, we describe the low-energy effective Hamiltonian for the RSW system, and illustrate the topological charges carried by such band structures. In Sec.~\ref{secfor}, we elaborate on the formalism to compute the response $\Gamma^{int} (\omega)$. Sec.~\ref{secnum} is devoted to the details of our numerical computations and results. In Sec~\ref{seciso}, we discuss the results for various topological semimetals with linear dispersions, in addition to the RSW case. We also point out the changes in the response for the cases of anisotropic dispersions.
In Sec.~\ref{sectilt}, we consider the effects of tilt and anisotropy, and elucidate the consequences. Finally, we conclude in Sec.~\ref{secend} with a summary and some outlook.
  
\section{The RSW Model} 
\label{secmodel}

In 3D, the low-energy single-particle RSW Hamiltonian, in the spirit of Eq.~\eqref{eq:kdots}, reads
\begin{align} 
	\label{eq:rsw}                                          
	\mathcal{H}_{\text{RSW}} = \hbar \, v_F \, \mathbf{k}\cdot \mathbf{J}\,, 
\end{align}
Here, $\mathbf{k}=\{k_x,k_y,k_z\}$ denotes the wave-vectors in the 3D BZ (measured with respect to the band-touching point), and $\mathbf{J}$ is given by the $4\times4$ pseudospin-3/2 matrices. These matrices follow cyclic commutation relations, and have the form 
\begin{align}
	& J_x= \begin{pmatrix} 
	0                   & \frac{\sqrt{3}}{2}   & 0                   & 0                    \\
	\frac{\sqrt{3}}{2}  & 0                    & 1                   & 0                    \\
	0                   & 1                   & 0                   & \frac{\sqrt{3}}{2}   \\
	0                   & 0                    & \frac{\sqrt{3}}{2}  & 0                    \\
	\end{pmatrix}, \nn
	& J_y= \begin{pmatrix} 
	0                   & -i\frac{\sqrt{3}}{2} & 0                   & 0                    \\
	i\frac{\sqrt{3}}{2} & 0                    & -i                  & 0                    \\
	0                   & i                    & 0                   & -i\frac{\sqrt{3}}{2} \\
	0                   & 0                    & i\frac{\sqrt{3}}{2} & 0                    \\
	\end{pmatrix}, \nn.
	& J_z= \begin{pmatrix} 
	\frac{3}{2}         & 0                    & 0                   & 0                    \\
	0                   & \frac{1}{2}          & 0                   & 0                    \\
	0                   & 0                    & -\frac{1}{2}        & 0                    \\
	0                   & 0                    & 0                   & -\frac{3}{2}         \\
	\end{pmatrix} .
\label{eqhamrsw}	
\end{align}
The energy eigenvalues are $\pm \frac{3\,k}{2},\pm \frac{k}{2}$, each associated with the respective Chern number $\mp 3, \mp 1$. Since the components of $\mathbf{J}$ are the generators of rotation, and they commute with the Hamiltonian, $\mathcal{H}_{\text{RSW}}$ is rotation-invariant.


\begin{figure*}
\centering
	\includegraphics[width = \textwidth]{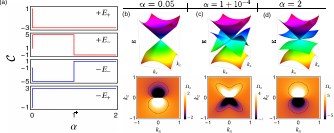}
\caption{\textbf{(a)} The Chern number $\mathcal{C}$ of each of the four bands of the Hamiltonian in Eq.~\eqref{eq:ham} is plotted against $\alpha$. The sum of the Chern numbers of the conduction bands $E_{\pm}$ represents the monopole charge, which changes at $\alpha=1$, indicating a topological phase transition. On the right, we consider three $\alpha$ values, and show the corresponding energy dispersions (denoted by $E$), and the contour-plots of the $z$-component of the ground-state Berry curvature $\mathbf \Omega$ (i.e., $\Omega_z$ for the LBB with dispersion $-E_+$). \textbf{(b)} For $\alpha=0$, the dispersion is two-fold degenerate. But as soon as we set $\alpha \neq 0$, the band degeneracy is lifted (except at $\mathbf{k}=0$). Here we show plots for $\alpha=0.05 $.
\textbf{(c)} There is a topological phase transition at $\alpha = 1$, and we show the behaviour at $\alpha = 1+10^{-4}$. 
\textbf{(d)} Finally, at $\alpha=2$, the Hamiltonian shows the RSW feature, with four bands crossing linearly at $\mathbf{k}=0$. The $\Omega_z$ lobes for $\alpha=1+10^{-4}$ or $\alpha = 2$ have signs opposite to those for $\alpha=0.05$, which demonstrates a change in the topological charge during the tuning of $\alpha$.} 
	\label{fig:1}
\end{figure*}

However, full rotational symmetry is not always guaranteed in real materials, and
$\mathrm{SO(3)}$ often reduces to the cubic rotation group $\mathrm{O_h}$. In such cases, the most generic Hamiltonian can be written as
\begin{align}
	\mathcal{H}_{\text{IF}}  
&=  \hbar \, v_F \, \mathbf{k} \cdot \left (\mathbf{V}+ \alpha \mathbf{U} \right ) 
\nn &=  \hbar \, v_F \sum \limits_{\mu} k_\mu \left (\frac{13 \,\alpha-14}{6}\, J_\mu
		+ \frac{8-4 \, \alpha}{6} \, J_\mu^3 \right ) ,	
\end{align}
where $\alpha $ is real. $ V_\mu=\frac{ -7\,J_\mu + 4 \,J_\mu^3} {3}$ and $ U_\mu
=\frac{ 13 \,J_\mu-4\,J^3_\mu } {6}$, related by 
$\tr \left[ V_\mu \,V_\nu \right ]=\tr\left [ U_\mu \,U_\nu\right ]
=4 \,\delta_{\mu \nu}$, and $\tr \left [V_\mu \,U_\nu \right ] =0$.
The second term is constructed from a set of
$4\times 4$ matrices that transforms as a vector under $\mathrm{O_h}$, and reduces full rotational symmetry to cubic rotation only. This problem was studied in detail by Isobe-Fu~\cite{isobe} in the context of $ S =3/2$ pseudospin fermions. Several candidate materials \cite{bradlyn,bradlyn3} have been proposed which realize the Isobe-Fu Hamiltonian. Examples include antipervoskite materials~\cite{isobe, hsieh} and transition metal silicides~\cite{tang}. Thus we use the Isobe-Fu type Hamiltonian to understand the topology and symmetry of the $ S=3/2$ fermions. When written explicitly in term of the $4\times 4$ matrices, this takes the form~\cite{igor, igor2}:
	\begin{align} 
		\label{eq:ham}
& \mathcal{H}_{\text{IF}} / \left( \hbar \, v_F \right)  
\nn &
= \begin{pmatrix} 
	k_z \left (1+\alpha \right ) & \frac{\sqrt{3}\left (k_x- i\,k_y \right ) \alpha} {2} & 0 &   
		\frac{ \left (k_x+i\, k_y \right ) \left (2-\alpha \right )} {2} \\
		\frac{\sqrt{3}\left (k_x + i\,k_y \right ) \alpha} {2} & k_z \left (-1+\alpha \right ) &  
		\frac{\left (k_x-i \,k_y \right) \left (2+\alpha \right )} {2} &  0 \\
		0 &  \frac{ \left (k_x+i\, k_y \right ) \left (2+\alpha \right )} {2} & 
		k_z \left (1-\alpha \right ) & 
		\frac{\sqrt{3} \left (k_x- i\,k_y \right ) \alpha}{2}  \\
		\frac{\left (k_x-i \,k_y \right ) \left (2 -\alpha \right )}{2} & 0 & 
		\frac{\sqrt{3}\left (k_x+ i\,k_y \right )\alpha} {2} &  
		-k_z \left (1+\alpha \right ) \\
		\end{pmatrix} .
	\end{align}
The eigenvalues are given by $\pm E_{\pm}$, where
\begin{widetext} 
\begin{align} 
E_{\pm}= \hbar \, v_F \sqrt{k^2 \left (1+\alpha^2 \right )              
\pm\alpha\,                                                        
		\sqrt{4 \left (k_x^4+k_y^4+k_z^4 \right )                          
		+ \left (3\,\alpha^2-4 \right )                                    
		\left (k_x^2 \, k_y^2+k_y^2 \,  k_z^2+k_x^2 \, k_z^2 \right )}}\,.
\end{align}
\end{widetext}
$E_{\pm}$ and $-E_{\pm}$ represent the dispersions of the conduction and valence bands, respectively. The most important feature of this Hamiltonian is that not only it captures the pseudospin-3/2 behaviour, but it also connects back to its other siblings like pseudospin-1/2 (or Weyl) system, through the tuning of the single parameter $\alpha$. The Chern number of each band for all the relevant ranges of $\alpha$ is illustrated in Fig.~\ref{fig:1}.

The parameter $\alpha$ essentially controls the interplay between the linear ($J_\mu$) and the cubic ($J_\mu^3$) terms.
For $\alpha=0$, we find $\mathcal{H}_{IF} \sim \mathbf{k}\cdot \mathbf{V}$, which gives isotropic dispersions.
In this limit, the Hamiltonian reduces to two pseudospin-1/2 Weyl cones of \textit{equal} chiralities, and is distinctive from the Dirac semimetals that decompose into a pair of Weyl points of \textit{opposite} chiralities. This can be better understood via a basis transformation $\mathbf{\Tilde{J}}=\tilde S\,
\mathbf{J}\, \tilde S^{-1} $ ($ {\tilde S}$ = unitary operator), where connetion to the WSM case or Pauli matrices becomes clearer as $\mathbf{V}  \rightarrow\mathbf{\Tilde{V}}  $ matrices take a more suggestive form, viz., $\mathbf{\Tilde{V}}= {\mathcal{I}}\otimes \boldsymbol{\sigma^{*}}$. The band dispersion in this limit is linear in $k$, and has a pair of degenerate bands with energy $\pm k$, as shown in Fig.~\ref{fig:1}(b) [note that Fig.~\ref{fig:1}(b) considers an $\alpha$ value slightly greater than zero, so that the degeneracy is slightly lifted].
The contour-plot of the $z$-component of the corresponding Berry curvature ($\mathbf{\Omega}$) of the LBB is shown below the dispersion plot, where a pair of lobes highlights the monopole structure ($\mathbf{\Omega}\sim  \mathbf{k}/|\mathbf{k}|^3$).
Since the Chern numbers are either $+1$ (for conduction bands) or $-1$ (for valence bands), the monopole charge turns out to be $2$. However, this topological charge changes from $2$ to $-4$ as $\alpha$ is tuned from $1^-$ to $1^+$ [see Figs.~\ref{fig:1}(b) and ~\ref{fig:1}(c)].
At $\alpha=2$, the Hamiltonian reduces to $\mathcal{H}_{\text{RSW}}$, which is rotation-invariant, and realizes the RSW fermions with linear dispersions [see Fig.~\ref{fig:1}(d)]. The corresponding Berry curvature landscape of the LBB is shown below the dispersion plot. Note that the inverted orientation of the lobes and increase in color bar intensity, in contrast with those for $\alpha=0.05$, confirm the changes in topological charge.

\section{Formalism} 
\label{secfor}

\begin{figure}[]
	\centering
	\includegraphics[width= 0.45 \textwidth]{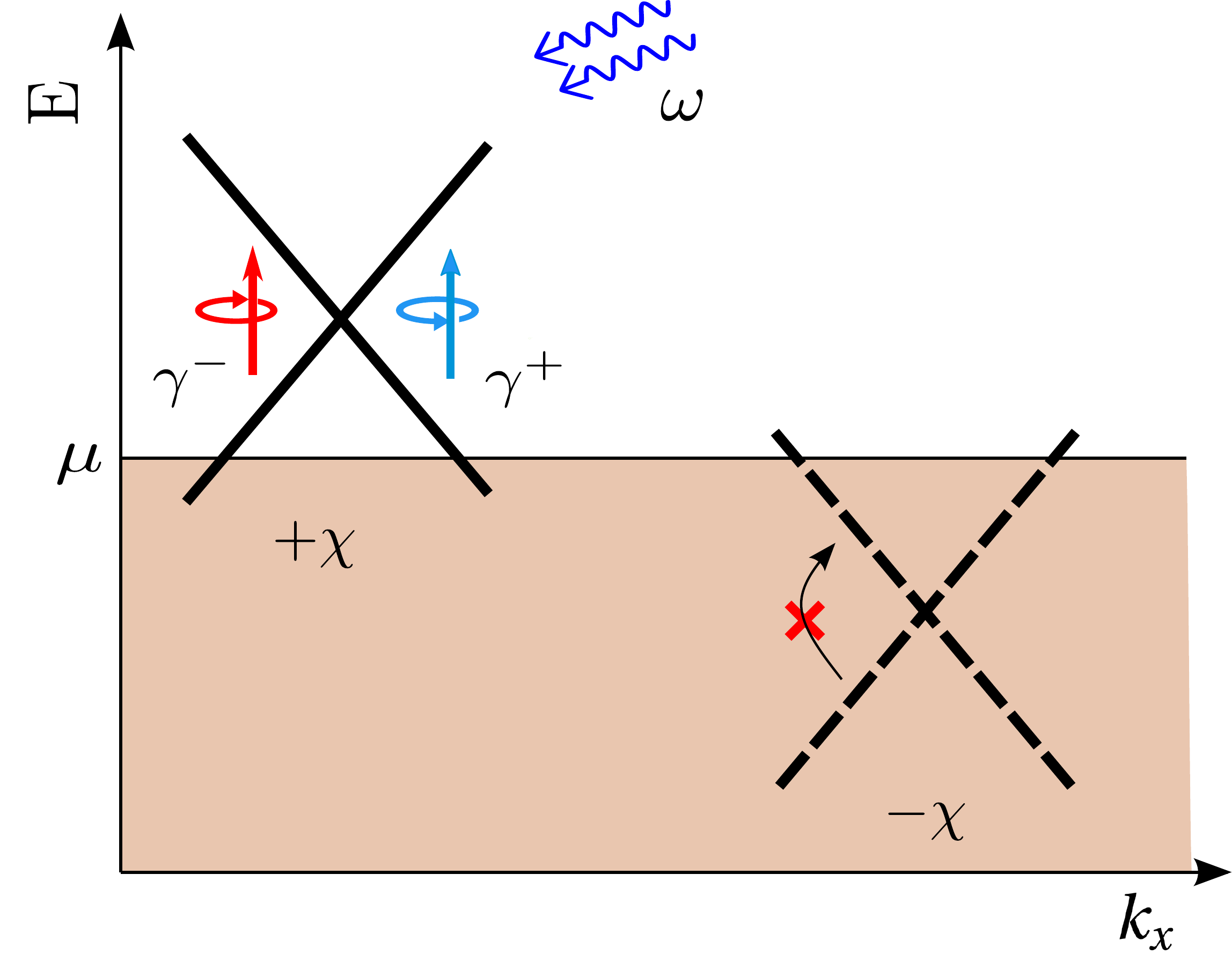}
	\caption{\label{fig:set-up}
Schematic band diagram representing the central idea behind the response $\Gamma^{int}(\omega)$: The dispersion $E$ is displayed against the momentum-component $k_x$, with $\mu$ as the Fermi level. Absorption of a probing photon of energy $ \hbar\, \omega$ leads to chiral transition rates $\gamma^{\pm}$. The net chiral depletion rate, multiplied by the difference in band velocities ($v$), yields a quantized quantity upon integration over the 3D BZ. Since a pair of Weyl nodes, with chiralities $\pm \chi$, lie at different energies, the Fermi level can be tuned to Pauli-block one of the nodes. In that situation, a single monopole contributes, yielding a quantized response.}
\end{figure}

Let us consider a 3D topological semimetal with the Hamiltonian $\mathcal{H}_0$, subjected to a monochromatic circularly polarized light of frequency $\omega$. We assume that the nodes of opposite chiralities lie at different energies, such that only the response of a single node matters (see Fig.~\ref{fig:set-up}).  
This circular drive, with two possible orientations (chiralities) denoted by $\pm$, can be written as a time-dependent perturbation $\mathcal{H'}_\pm (t)= 2\,\mathcal{E} \left [-\sin(\omega \,t) \,\hat{\mu} \pm \cos(\omega\, t)\,\hat{\nu} \right ]$, where $\mathcal{E}$ is the amplitude of the circular drive, and has the units of force \footnote{Experimentally, such a perturbation can be achieved in an optical lattice with ultracold fermionic atoms, via a frequency modulation of the lattice beams~\cite{asteria}}. Furthermore, $\hat{\mu}$ and $\hat{\nu}$ denote the mutually perpendicular position vector operators, in the plane of polarization of the applied drive.

The perturbation results in a modulation of the momenta as
\begin{align}
& k_{\mu} \rightarrow k_{\mu} + \frac{2 \, \mathcal{E}}{\hbar \, \omega} 
 \cos(\omega \, t) \,,
\quad k_{\nu} \rightarrow k_{\nu} \pm \frac{2 \, \mathcal{E}}{\hbar \, \omega} \, \sin(\omega \, t)\,.
\end{align}
First, we perform a unitary transformation to a frame where the Hamiltonian is translation- invariant~\cite{goldman_dir,ozawa}. Then, assuming that the amplitude of the drive is small (i.e., $\mathcal{E} / \left(\hbar \, \omega \right) \ll 1 $), we expand the Hamiltonian in a Taylor expansion, and restrict ourselves to the first-order terms, to obtain the form
\begin{align} 
\label{eq:pert}
	{\mathcal{H}}_{\pm}(t)\approx \mathcal{H}_0 + 
\frac{2\, \mathcal{E}} {\hbar \,\omega} 
	\left [ \cos(\omega \,t) \, \frac{\partial \mathcal{H}_0}{\partial k_{\mu}}                  
	\pm  \sin(\omega \,t)\, \frac{\partial \mathcal{H}_0}{\partial k_{\nu}} \right ].             
\end{align}
For a system with non-degenerate bands, we populate an initial state, usually the LBB (labeled by $n$), and compute the transition rate to an excited state (labeled by $m$) for long times. This transition (or depletion) rate is given by Fermi's golden rule~\cite{sakurai} in the time-dependent perturbation theory, and can be written as (see the Appendix)
\begin{align}
\label{eqgampm}
& \gamma_{mn}^{\pm}(k_\mu,k_\nu,\omega) 
\nn &=              
 \frac{ 2\,\pi} {\hbar}
	 \left( \frac{\mathcal{E}}{\hbar \, \omega} \right)^2                                                    
	\Big| \mathcal{P}_{mn}^ \mu \mp i \,\mathcal{P}_{mn}^ \nu \Big|^2 
	\delta(E_{mn}- \hbar \,\omega) \,.                                         
\end{align}
Here, $\mathcal{P}_{mn}^ \mu=\braket{m|\frac{\partial \mathcal{H}_0}{\partial k_{\mu}}|n}$ is the $\mu$-component of the optical matrix element between $m^{\text{th}}$ and $n^{\text{th}}$ bands, and the energy difference is $E_{mn} = E_m-E_n$.
To check that the dimensions match correctly on both sides of the equation, we note that on the RHS, $\delta(E_{mn}- \hbar \,\omega) / \hbar$ has the dimension of inverse time and the rest of the expression is dimensionless. Thus, $\gamma_{mn}^{\pm}(k_\mu,k_\nu,\omega) $ gives the total number of quasiparticles scattered from the LBB per unit time, which can be measured experimentally. Due to the delta function, a transition involving the $m^{\text{th}}$ and the $n^{\text{th}}$ bands yields a finite response when an incident photon has an energy exactly equal to $E_{mn}$. However, this precise energy matching is only an idealization -- in reality, there exists a small energy window (due to the uncertainty relation) within which photons of slightly different frequencies will get absorbed. Thus, the delta function is often approximated as a Lorentzian with a small broadening.
We further point out that the interband transitions create a Rabi oscillation at each $\mathbf{k}$, contributing to a linear growth of $\gamma_{mn}^{\pm}$. But as pointed out in Ref.~\cite{goldman_dir}, this effect is suppressed in a realistic material with disorder. 

We now define the local differential rate by
\begin{align}
\label{eq11}
\Delta \gamma_{mn}^{\mu \nu} (\mathbf k,\omega)& = 
\frac{ \gamma_{mn}^{+}(k_\mu,k_\nu,\omega)  - \gamma_{mn}^{-}(k_\mu,k_\nu,\omega) } 
{2} \, ,
\end{align}
and multiply it with the difference in band velocities, $v_{mn}^{\lambda}
=\partial E_{mn} / \partial k_{\lambda}$, along the direction $\lambda$ which is perpendicular to the plane of the drive. This defines the quantity (more details are provided in the Appendix)
\begin{align}
\label{eq12}
& \Gamma(\mathbf{k}, \, \omega)                                       
= \sum_{n\in \mathbf{FB}}  \sum_{m\neq n}  
 \sum_{\mu,\nu,\lambda} \frac{\epsilon_{\lambda \mu \nu}} {2}\,
 v_{mn}^{\lambda} \, \Delta \gamma_{mn}^{\mu \nu}	\nn &                                           
= \frac{2\,\pi \,i\, {\mathcal{E}}^2} {\hbar^3 \, \omega^2}
  \sum_{n\in \mathbf{FB}} \sum \limits_{m\neq n} 
  \sum_{\mu,\nu,\lambda} 
  \epsilon_{\mu \nu \lambda} \, v_{mn}^\lambda \, \mathcal{P}_{mn}^ \mu \,
  \mathcal{P}_{nm}^\nu\,\delta(E_{mn}- \hbar \, \omega)\,,
\end{align}
where $\epsilon_{\mu \nu \lambda}$ denotes the cyclic permutation (Levi-Civita symbol), and the index $n$ spans over the initially filled bands ($\mathbf{FB}$s).
Finally, integrating $\Gamma(\omega)$ over the BZ leads to a net depletion rate
\begin{align} 
\label{eqgamma}
\Gamma^{int}(\omega) =                                                                      
\int \, \left[d \mathbf{k} \right] \, \Gamma (\mathbf{k}, \omega) \,,
\quad \left[d \mathbf{k} \right] = \frac{d k_x \, d k_y \, d k_z }  {(2\,\pi)^3}\,,
\end{align}
which depends on the frequency of the incident photons, and is the DIR for 3D systems explained in the introduction.
This response can be measured experimentally for WSMs without any mirror symmetry. Materials like CoSi, RhSi, AlPt have space group $\mathrm{P2_1 3 \, (SG \, 198)}$, and have a chiral tetrahedral symmetry. Lack of mirror planes in these structurally chiral materials causes the Weyl nodes to lie at different energies~\cite{weber}, such that the Fermi level can be tuned to Pauli-block one of the nodes, rendering it inert, so that a single node contributes to the optical response (cf. Fig.~\ref{fig:set-up}). The Pauli blockade is an important factor to prevent cancellations from nodes carrying opposite topological charges. Considering energy resolution, $\mathrm{RhSi}$ hosts a fourfold point at $\Gamma$, with the lowest two topological bands having a span of $40$ meV. This is well within the range of THz spectroscopy ($\sim 0.1-10$ THz), and coherent meV photons can be used to probe this system.

We now discuss the situation when $\Gamma^{int}(\omega)$ is quantized.
Let us take a system of non-degenerate bands, where the band index is denoted by integers. We initially populate the ground state (LBB) of this system at time $t=0$. Considering transitions that are only taking place from the LBB ($n=0$) to higher excited states ($m=1,2,3...$), we can write
\begin{align}
& \Gamma^{int} (\omega)  
\nn &= 
\frac{2\,\pi \,i\, {\mathcal{E}}^2} {\hbar^3 \, \omega^2}
	\int \left[d \mathbf{k} \right] 
\sum_{\mu,\nu,\lambda} \epsilon_{\mu \nu \lambda} \,
	\Big[ v_{10}^\lambda\, \mathcal{P}_{10}^\mu\, \mathcal{P}_{01}^{\nu} \, 
	\delta (E_{10}- \hbar \, \omega )
	\nn & \hspace{ 2 cm}
	+ v_{20}^\lambda \,\mathcal{P}_{20}^\mu\, \mathcal{P}_{02}^{\nu}
	\,\delta(E_{20}- \hbar \, \omega)+ \ldots \Big]\,.
\end{align}
We exploit the presence of the delta function and keep $\omega$ low (in particular, compared to the next to lowest energy-difference $E_{20} $), which makes every other term vanish except the first. From the definition, the Berry curvature of the LLB is
\begin{align}
& \Omega^{0}_\lambda = i\,
\sum \limits_{\mu,\nu} \epsilon_{\mu \nu \lambda} \sum_{m\neq 0} 
	\frac{\mathcal{P}_{m0}^\mu \, \mathcal{P}_{0m}^{\nu} }{E_{m0}^2}\,.
\end{align}
It is convenient to rewrite the above identity as
\begin{align}
& i\, \sum \limits_{\mu,\nu}	\epsilon_{\mu \nu \lambda}\, 
\mathcal{P}_{10}^ \mu\,\mathcal{P}_{01}^ \nu
\nn & =	E_{10}^2 \left [  \Omega^{0}_\lambda                                            
- i \, \sum \limits_{\mu,\nu} \epsilon_{\mu \nu \lambda}\sum_{m\neq 0,1}                             
\frac{\mathcal{P}_{m0}^ \mu \, 
	\mathcal{P}_{0m}^{\nu} } {E_{m0}^2} \right ],         
\end{align}
which immediately leads to
\begin{align}
&	\Gamma^{int} (\omega) 
\nn & \simeq 
\frac{2\,\pi \, {\mathcal{E}}^2} {\hbar^3 \, \omega^2}                                                   
\sum \limits_{\lambda}	
\int \left[d \mathbf{k} \right]  
\left(  \Omega^{0}_\lambda 
- i \,\sum \limits_{\mu,\nu} \epsilon_{\mu \nu \lambda}\sum_{m\neq 0,1} 
\frac{\mathcal{P}_{m0}^\mu \, \mathcal{P}_{0m}^\nu } {E_{m0}^2}
\right) 
\nn & 
\hspace{ 3 cm} \times v_{10}^\lambda\,
E_{10}^2 \,\delta (E_{10}- \hbar \, \omega )\,.
\label{eqapp}
\end{align}

Eq.~\eqref{eqapp} involves a 3D integral in general, but it can be reduced to a surface integral for isotropic systems. This is better understood if we switch to the spherical polar coordinates ($k, \theta, \phi$), and consider the case of an isotropic system, such that generically the dispersion $E_{10}$ is a function of $k$ only (i.e., with no dependence on the angular coordinates). This implies that
$\mathbf v_{10}  =  \frac{1}{\hbar} \frac{d E_{10}} {dk}\, {\mathbf{ \hat k}}$. 
With this prescription, the first term on the RHS of Eq.~\eqref{eqapp} can be reduced to
\begin{align} 
\label{eq:chern_appx}
& \frac{2 \, \pi  \, \mathcal{E}^2}{\hbar} 
\int \frac{dk\,d\theta\,d\phi\, \sin \theta \,k^2}
{(2\,\pi)^3}                                  
	\left( \frac{\,E_{10}^2 } {\hbar^2 \, \omega^2} \right) \,
\delta( E_{10}- \hbar \, \omega) \,{\mathbf v}_{10} \cdot \mathbf{\Omega}^{0} 
\nn & =  \left( \frac{\mathcal{E}}{\hbar} \right)^2  
\int dE_{10}\, \frac{\,E_{10}^2 } {\hbar^2 \, \omega^2} \, \delta( E_{10}- \hbar \, \omega)\, 
 \nn  & \hspace{ 2.5 cm} \times 
 \frac{1}    { (2\,\pi)^2} 
\int \,d\theta\,d\phi\, \sin \theta \,k^2 \, {\mathbf{ \hat k}} \cdot \mathbf \Omega^{0} 
\Big \vert_{ E_{10}(k) =\hbar \,\omega } \nn
\nn & =  \left( \frac{\mathcal{E}}{\hbar} \right)^2   
\frac{1}{(2\,\pi)^2} 
\int_S \,d\mathbf S \cdot \mathbf \Omega^{0} 
 = \left( \frac{\mathcal{E}}{\hbar} \right)^2 \frac{\mathcal{C}_0}{2 \, \pi}\,,
\end{align}
where $\mathcal{C}_0$ is the Chern number of the LBB. In the intermediate steps,
$ d\mathbf S = \tilde k^2 \,\sin \theta \,d\theta\,d\phi\,{\mathbf{ \hat k}} $ denotes an infinitesimal area vector on the surface of the sphere $S$ centred at the origin (i.e., $k=0$), whose radius $\tilde k $ is obtained by solving for $k$ in $E_{10} (k) =\hbar \,\omega$.
The second term of Eq.~\eqref{eqapp} can be treated in an identical manner. This leads to the central result that
\begin{align}
\label{eq15:dir}
	  & \Gamma^{int}(\omega) \nn & = 
\frac{(\mathcal{E} / \, \hbar )^2}{2 \, \pi} \,  
	\left[ \mathcal{C}_0 
- \frac{i} {2\,\pi} \int_S d\mathbf S \cdot  \sum_{m\neq 0,1}  
	\frac{
	\boldsymbol{ \mathcal{P}}_{m0} \times \boldsymbol{\mathcal{P}}_{0m} }
	{E_{m0}^2} \right] .
\end{align}
Interestingly, in the quantized limit (i.e., when the first term on the RHS dominates), $\Gamma^{int}(\omega)$ is independent of the microscopic parameters of the Hamiltonian, and is controlled only by the external drive strength $\mathcal{E}$. Experimentally, $\mathcal{E}$ is set by the ratio of the recoil energy $E_r$ to the lattice spacing $a$. For example, in an optical lattice setup~\cite{asteria} with $\mathrm{{}^{40}K}$ atoms, $a=410$ nm, $E_r / h = 4.41$ kHz, and $\mathcal{E}$ is equal to $0.6-1.2\%$ of $E_r/a$.

It is crucial to note that we could arrive at this quantized result for an isotropic system. For an anisotropic system, with the energy and other physical quantities depending on the angular coordinates, this simplification is not possible.
The multi-band correction terms, captured by $\Delta_{m>1} \equiv 
- \frac{ i\, {\mathcal{E}}^2} {4\,\pi^2 \, \hbar^2} 
 \int_S d\mathbf S \cdot  \frac{
	\boldsymbol{ \mathcal{P}}_{m0} \times \boldsymbol{\mathcal{P}}_{0m} }
{E_{m0}^2}$, are present for a system with more than two bands, and the quantization condition emerges in the limit when we can neglect these correction terms.
For a system with linear dispersions, $E_{mn}\sim k$, which makes it straightforward to infer
that $E_{m0} \sim \hbar \, \omega $ due to the delta function integral, suggesting that the correction terms decay quadratically with frequency.

To elucidate the quantization features of $\Gamma^{int}$, we consider three kinds of topological semimetal, viz., WSM, pseudospin-1, and the four-band model defined in Eq.~\eqref{eq:ham}, and compute the normalized differential current as a function of the drive frequency $\omega$. Note that the linearized $\mathbf{k} \cdot \mathbf {p}$ Hamiltonian for pseudospin-1 fermions has the form
\begin{align}
	\mathcal{H}_1(\mathbf  k) = 
\hbar \, v_F \, \mathbf{k}\cdot \mathbf L\,, 
\end{align}
where $\mathbf L $ represents the vector spin-1 operator with the three components
\begin{align}
	& L_x = \frac {1} {\sqrt{2}}
	\begin{pmatrix}
	0 & 1 & 0 \\1&0&1\\0&1&0
	\end{pmatrix},\,
	L_y =\frac{1}{\sqrt{2}}
	\begin{pmatrix}
	0&- i  &0\\  i  &0&-  i \\
	0&  i  &0
	\end{pmatrix},\nn
	& L_z =
	\begin{pmatrix}
	1 & 0 & 0 \\0&0&0\\0&0&-1
	\end{pmatrix}.
\end{align}

\section{Numerical Results}
\label{secnum}

In our numerical calculations, we have used Eq~\eqref{eqgamma} (without any analytical approximation) to compute the response in the range $\omega \in [0,3]$, with a scanning interval of $\Delta \omega=0.1$. The Dirac delta functions in the equation are incorporated  by 
$\delta(x)=\frac{1}{\pi}  \, \lim \limits_{\varepsilon\rightarrow 0} \frac{\varepsilon}{x^2+\varepsilon^2}$, where $\varepsilon$ controls the broadening of the function, and should be kept small (we have set it to $10^{-4}$ in our numerics). Due to the radial symmetry of the systems, we have used the spherical polar coordinates, and evaluated the integrals via the ``global adaptive'' method (with a precision goal of $10^{-5}$) of the \textit{Mathematica} software. Note that the upper limit of the $k$-integral is $\infty$ in a continuum model. In our numerics, we have set this upper limit to 100, which ensures convergence of the results.

\subsection{Untilted Isotropic Cone} 
\label{seciso}


\begin{figure}[] 
\centering
\includegraphics[width = \columnwidth]{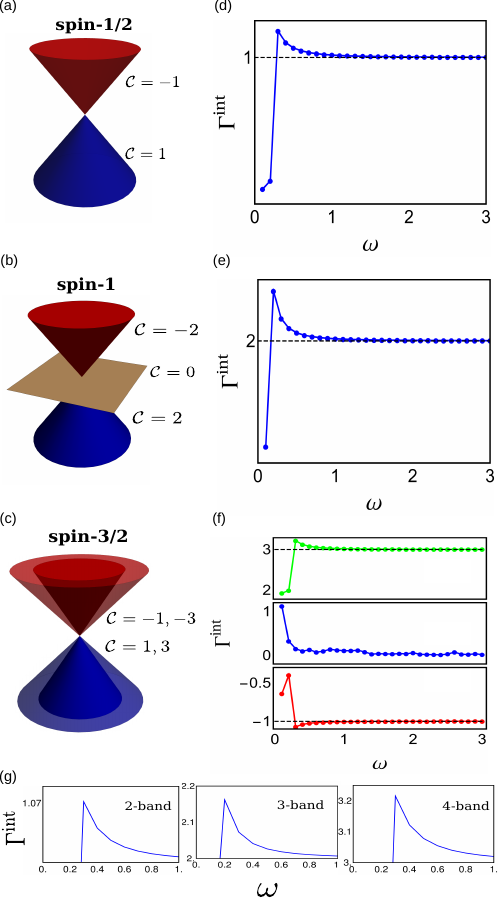}
\caption{\label{fig:3}
\textbf{(Left panel)} The energy bands of the \textbf{(a)} WSM, \textbf{(b)} pseudospin-1 fermions, and \textbf{(c)} four-band model of Eq.~\eqref{eq:rsw}  are shown, along with the respective Chern numbers. \textbf{(Right panel)} Numerical plots of $\Gamma^{int} $ as a function of $\omega$ are shown for comparison. \textbf{(d, e)} For each of the WSM and pseudospin-1 cases, $\Gamma^{int} $ versus $\omega$ is found to be quantized (above a critical value $\omega_c$) at moderate to high frequencies, and is equal to the LBB Chern number. \textbf{(f)} This plot shows $\Gamma^{int}$ versus $\omega$ for the four-band model of Eq.~\eqref{eq:ham}, at three distinct values of $\alpha$. While $\Gamma^{int} (\omega>\omega_c)$ correctly infers the Chern numbers for $\alpha = 10^{-3}$ (red)  and $\alpha = 2$ (green), the response decays to zero for $\alpha = 1 + 10^{-4}$ (blue). \textbf{(Bottom panel) (g)} The deviation of $\Gamma^{int}$ from the expected quantized value is shown to highlight the effect of higher-band inclusions. The ``4-band'' here refers to the isotropic RSW case. Each additional band increases the peak response in accordance with the theory. In all the plots, $\Gamma^{int} $ is in units of $\frac{ \mathcal{E}^2}{2\, \pi \, \hbar^2}$.}
\end{figure}

As seen from Fig.~\ref{fig:3}(d) and \ref{fig:3}(e), the quantized response characteristics exist for both the pseudospin-1/2 and pseudospin-1 cases. At low-frequency ranges of $\omega<<1$, $\Gamma^{int}$ yields a small response, which dramatically increases as $\omega$ crosses a threshold value $\omega_c$.
This threshold value arises in experimental measurements due to the limitations
in instrumental accuracy. In our numerics, it depends on the $k$-mesh resolution. In the regime $\omega>\omega_c$, the response decays with $\omega$ in a quadratic fashion. As $\omega$ increases, $\Gamma^{int}$ saturates to a constant number, determined by the LBB Chern number. 
This quantization is accurate up to two decimal places in our numerics, and prevails beyond $\omega=3$. 

However, our results for the four-band model of Eq.~\eqref{eq:ham} reveal that the quantization observed for the above cases is not universal, as the quantization disappears depending on the value $\alpha$. In Fig.~\ref{fig:3}(f), $\Gamma^{int}$ is quantized in the range $\omega>\omega_c$, and correctly probes the Chern numbers for $\alpha= 10^{-3}$ and $\alpha = 2$. For example, in the RSW limit of $\alpha=2$, we find that $\Gamma^{int}$ saturates to $3$, which is consistent with the CPGE results for RhSi at the fourfold point~\cite{chang}. But,
such a quantization is absent in the case of $\alpha=1+10^{-4}$, as the response decays to zero. We further point out that the absence of quantization prevails for any $\alpha$ value, except
the special points of $\alpha=0^+, 2$ (where the energy dispersions depend linearly in $k$). This seems to be related to the fact that the energy dispersion is not isotropic for $\alpha\neq 0,2$. 
The anisotropy makes the integral in Eq.~\eqref{eqgamma} frequency-dependent for all $\omega$ values, such that it cannot be reduced to a quantized form (modulo the correction terms $\Delta_{m}$) of Eq.~\eqref{eq15:dir}. As the probing frequency increases and goes beyond the scale of the band gaps, the response function quickly decays to zero due to the delta function, which is consistent with the $\alpha=1 +10^{-4}$ result.

Lastly, we want to focus on the multi-band correction terms $\Delta_{m>1}$ in the analytical expression of Eq.~\eqref{eq15:dir}, which are crucial for higher-pseudospin systems. As shown before, the key to quantization in Eq.~\eqref{eq15:dir} comes from $\hbar \, \omega\sim E_{01}$, which generally requires the probing frequency to be small. At the same time, the presence of multiple bands adds a frequency-dependent term, that is coincidentally strongest in a similar frequency range. This can deviate the result from the expected Chern number. We provide the plots of $\Gamma^{int}$ in Fig.~\ref{fig:3}(g) for lower frequency ranges to showcase this effect. Since $\Gamma^{int}\sim0$ for $\omega<\omega_c$, an insight into $\Delta_{m}(\omega)$ can be gained by looking at the peak value $\Gamma_{max}^{int}(\omega\rightarrow\omega_c^{+})$ of $|\Gamma^{int}|$, just as $\omega$ goes above $\omega_c$. Evidently, this correction is very low ($\sim 0.07$) for the two-band WSM. The fact that it is not exactly zero for the Weyl case can be explained when we note that we have implemented the Dirac delta function via numerical approximation. Thus, when $ \hbar\, \omega\rightarrow E_k+\delta E_k$, the delta function $\delta(E_k-\hbar\, \omega)$ still picks up a small non-zero contribution, which causes this overestimation. However, as we consider the pseudospin-1 system, and have one more band compared to the two-band scenario, $\Gamma_{max}^{int}$ increases by an order of magnitude (from $\sim 0.07$ to $\sim 0.2$). The same effect can be seen for the four-band case as well, supporting the calculations done in this paper. 
Our analytical calculations, followed by in-depth numerics, test the idea of using circular dichroism to detect topological invariants, particularly in 3D, complementing existing optics-based techniques. Moreover, our results strengthen previous works on DIR, in the sense that it shows  how monitoring DIR can be a viable scheme to obtain a quantized response, at least for two-band systems where many-band correction terms are negligible.

\begin{figure}[] 
	\centering
	\includegraphics[width=0.4\columnwidth]{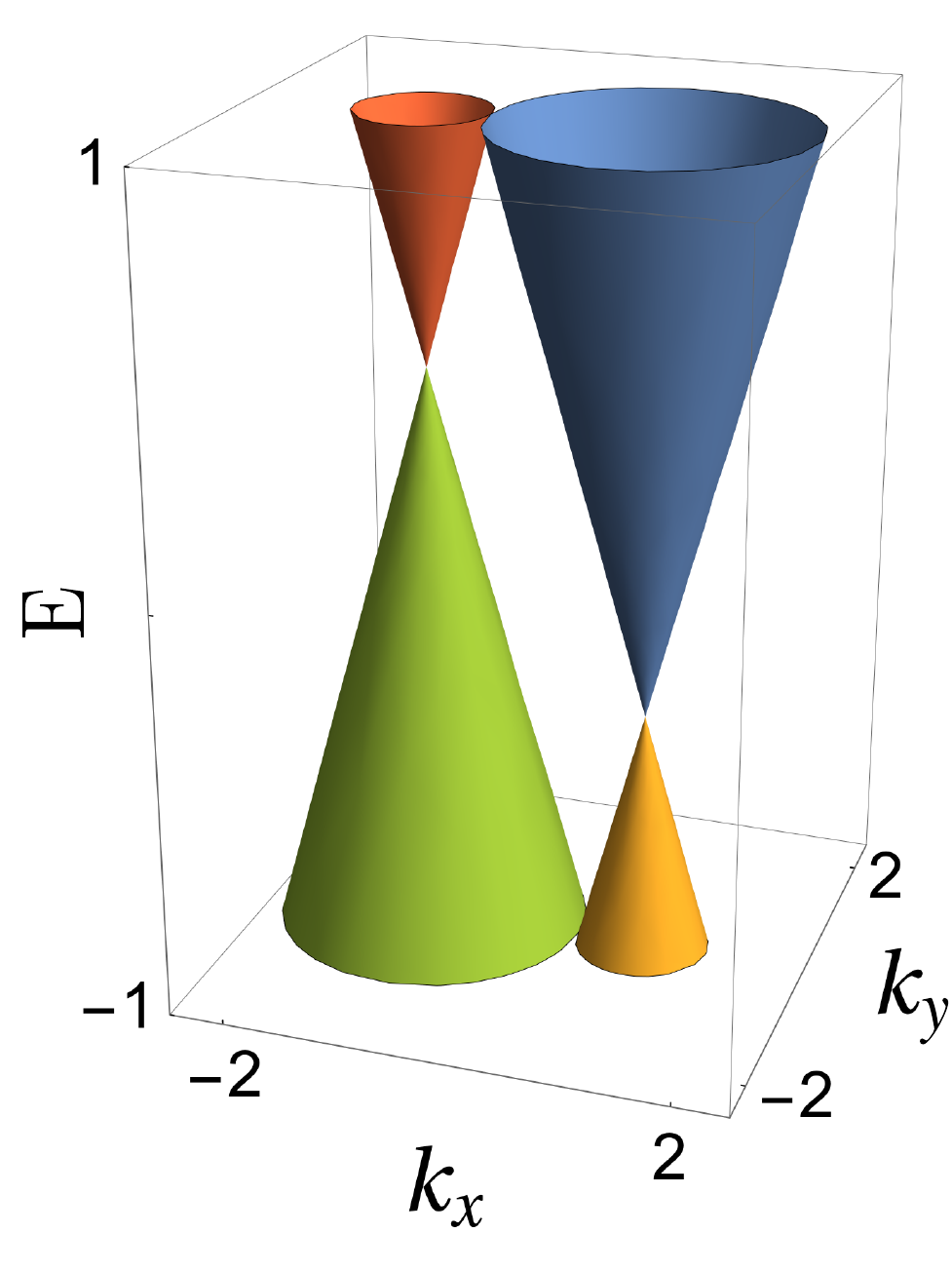}
\caption{\label{fig:4}
Energy spectra of the pair of Weyl nodes obtained by adding a perturbation
$\left( \delta \,s_x\, \tau_y  + b \, s_z \, \tau_z \right)$ to the 3D Dirac semimetal of Eq.~\eqref{eqdirac}, thereby splitting the Dirac cone. Here, the parameter $\delta$ tunes the distance between nodes in the momentum space, and $b$ separates the nodes in energy. In the figure, we have set $\delta=1$ and $b=0.4$.}
\end{figure}

So far, we have shown results for systems carrying nonzero topological charges only, our methodology should obtain zero response for cases with zero topological charges. 
Here we present a trivial situation, for which the response is zero. In particular, we consider the 3D Dirac semimetal with the Hamiltonian~\cite{yang}
\begin{align}
\label{eqdirac}
	\mathcal{H}_{\text{dirac}} =
\hbar \, v_F \, \left( k_x \,\sigma_y \, \tau_x  - k_y \,\sigma_x \,\tau_x  
+  k_z \,\tau_z \right),
\end{align}
where $\boldsymbol{\sigma}$ and $\boldsymbol{\tau}$ denote the vectors of the three Pauli matrices acting on two different symmetry-operator spaces. The system contains doubly-degenerate cones with dispersions $  \pm k$. Adding the perturbation $\delta \, \sigma_x \, \tau_y$ splits a Dirac node into two distinct Weyl nodes, separated in the $k_x$-momentum direction. These have dispersions $\pm E_{\text{weyl},s}$, where $E_{\text{weyl},s} =  \hbar \, v_F \, \sqrt{ \left [ k_x + (-1)^s\, \delta \right ]^2+k_y^2+k_z^2}$, and $s=1,2$ labels the two Weyl nodes. Notice that the inversion symmetry is preserved, and the response integral for $\Gamma^{int}$ now contains two Weyl nodes of opposite chiralities at the same energy. This leads to a trivial zero response, as in Eq.~\eqref{eq15:dir}, we have now two ${\mathcal C}_0$'s with values $\pm1$. A non-trivial response is obtained when the integral contains one or more Weyl nodes with the same chirality [e.g., the $\alpha =0 $ case of the Isobe-Fu Hamiltonian of Eq.~\eqref{eq:ham}].
We can further add a term $b \, s_z \, \tau_z$ to break the inversion symmetry, which shifts the Weyl nodes to different energies [cf. Fig.~\ref{fig:4}], such that the dispersions are now $ E_{\text{weyl},s}^{\pm}  / \left( \hbar \, v_F  \right)
=  (-1)^s\,b  \pm \sqrt{ [ k_x + (-1)^s\, \delta ] ^2+ k_y^2+ k_z^2}$.
Experimentally, such Dirac to Weyl transitions can be achieved in 3D sonic crystals with the introduction of chiral hoppings~\cite{xie}.
If these nodes are reasonably separated in energy, we can tune the chemical potential such that it lies in the conduction band (i.e. $ E_{\text{weyl},1}^+$) of the Weyl node with the lower energy, while cutting the valence band (i.e. $ E_{\text{weyl},2}^-$) of the other Weyl node.
With this Fermi level tuning, only the second Weyl node contributes to the currents,
and the problem reduces to the $2\times 2$ Hamiltonian
$\mathcal{H}_W =  \hbar \, v_F \, \chi\,\boldsymbol{k}. \boldsymbol{\sigma}$, after linearizing the momentum dependence around node $2$ (where $\chi$ denotes the chirality of the node in consideration). We have already shown this to yield a quantized response [cf. Fig.~\ref{fig:3}(a), (d), (g)].

\subsection{Effects of Tilt and Anisotropy} 
\label{sectilt}

In real materials, the dispersion around the band-touching point is not perfect. Depending on the space group symmetries, one may encounter possible factors, especially tilt and anisotropy that are ubiquitous in semimetals. We use this section to discuss how such factors, when present, affect the DIR. Tilt appears due to the absence of particle-hole symmetry, and is constrained by the point-group symmetries of condensed matter physics. On the other hand, anisotropy may arise from imperfections in materials, leading to a deviation from the isotropic band-dispersions. Particularly, recent experimental Fermi surface data of AlPt~\cite{schroter} suggest that the role of such effects can be a prominent factor. Therefore, we consider an effective RSW Hamiltonian with tilt and anisotropy (in the spirit of Ref.~\cite{trescher} for the Weyl case), given by
\begin{align}
	\mathcal{H}_{mod} =  \left ( \hbar\, v_t\,k_z -m_{\chi} \right )\mathcal{I}_4 
+ \hbar\, v_F
	\sum_{\mu,\nu} \eta_{\mu \nu}\,k_\mu \, J_\nu \,.              
	\label{eqaniso}                                                       
\end{align}
Here, $v_t$ represents the tilt velocity along the $z$-axis, $\mathcal{I}_4$ denote $4\times4$ identity matrix, $v_F$ is the Fermi velocity, $\chi= \text{sgn}\left [\text{Det}(\eta) \right ]$ determines the chirality, and $m_{\chi}$ sets the energy at which the node of chirality $\chi$ is located. We set $m_{\chi}$ to zero at one of the nodes for simplicity, and consider that one for further calculations. Being proportional to $\mathcal{I}_4$, the first term is trivial, and is responsible for breaking inversion symmetry. The second term in $\mathcal{H}_{mod}$ contains the $3\times 3$ anisotropy matrix $\eta $ (with components $\eta_{\mu \nu}$). It is easy to check that $\eta_{\mu \nu}=\delta_{\mu \nu}$ for the isotropic case. Note that every possible anisotropy of a cone (up to linear order in $k$) can be modelled by an upper triangular form for $ \eta $. Thus, we write the anisotropy matrix, without any loss of generality, as
\begin{align}
	\eta =  \begin{pmatrix} 
	\eta_{xx} & \eta_{xy} & \eta_{xz} \\
	0         & \eta_{yy} & \eta_{yz} \\
	0         & 0         & \eta_{zz} \\
	\end{pmatrix}.
\end{align}
We set the diagonal terms equal to one (i.e., $\eta_{xx}= \eta_{yy}= \eta_{zz}= 1 $), and the off-diagonal terms to $r$ (i.e., $ \eta_{xy}= \eta_{xz}= \eta_{yz}= r $). In this way, we consider a particular realization of anisotropy caused only by off-diagonal terms of equal values.

The linear energy dispersion of the untilted and isotropic RSW gets modified to
\begin{widetext}
\begin{align}
\label{eqdisaniso}
E_{\pm}^{p}   / \left( \hbar \, v_F  \right)\,                                               
		=  \zeta \, k_z \pm p \,                                   
		\sqrt{k^2 + 2\,r                                           
		\left (k_x \,k_y+k_y \,k_z+k_x \,k_z \right )              
		+r^2 \left (2\,k_{x}^2+2\,k_x \,k_y+ k_{y}^{2}\right )}\,, 
		\quad  p=\frac{1}{2},\, \frac{3}{2}\,.                     
	\end{align}
\end{widetext}
We set $ \zeta = v_t/v_F$, so that the eigenspace is characterized by two parameters, namely, $\zeta$ and $r$. In the following, we take $r$ and $\zeta$ to be zero alternatively to understand the individual effects of tilt and anisotropy.

\begin{figure}[]
	\centering
\includegraphics[width=\columnwidth]{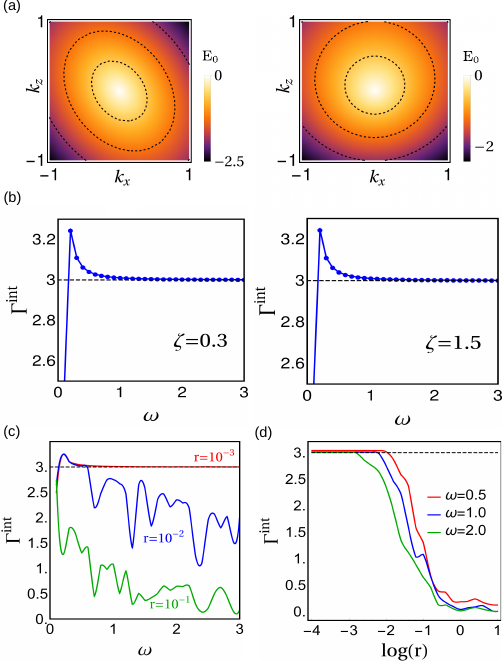}
\caption{\label{fig:5}
RSW semimetal in the presence of tilt and anisotropy [cf. Eq.~\eqref{eqdisaniso}]: 	
\textbf{(a)} The contour-plot of the dispersion of the LBB (denoted by $E_0$) is shown in the $k_x$-$k_z$-plane, in the presence of tilt (left), or anisotropy (right). With $r=0$ and $\zeta =0.3$ in the left panel, we see that $\zeta $ controls the tilt along $z$-axis. In the right panel, we have $r=0.3$ and $\zeta =0$, showing that a finite value of $r$ amounts to ``squeezing'' of the dispersion in certain directions. \textbf{(b)} Since tilt does not affect $\Gamma^{int}$, the responses in type-I ($r=0, \zeta=0.3$) and type-II ($r=0, \zeta=1.5$) phases are similar. \textbf{(c)} For $r=10^{-3}$ and $\zeta=0$, $\Gamma^{int}$ exhibits a quantized plateau at all frequencies above a critical value $\omega_c$. A considerable deviation from the plateau is seen at moderate to high frequencies if we set $r=10^{-2}$, followed by a non-quantized response at $r=10^{-1}$. \textbf{(d)} Here we show $\Gamma^{int}$ against $\log(r)$ for three different photon frequencies. In all the plots, $\Gamma^{int} $ is in units of $\frac{ \mathcal{E}^2}{2\, \pi \, \hbar^2}$.}
\end{figure}

First, we set $r=0$ and $\zeta \neq 0$, in order to consider the case of isotropic tilted cones. The presence of a finite tilt can be inferred from the asymmetric dispersion of the LBB, as shown in Fig.~\ref{fig:5}(a). We have shown $\Gamma^{int}$ for two different tilt phases in Fig.~\ref{fig:5}(b). It is clear that the response is independent of any tilt value. This is because, tilt, being coupled to the identity matrix, does not affect the eigenspinors -- hence, the optical matrices $\mathbf{\mathcal{P}^\mu}$ in Eq.~\eqref{eq12} do not change. Additionally, the difference of band-velocities $\mathbf v_{mn}$ also remain invariant under tilt. By virtue of these combined facts, Eq.~\eqref{eqapp} can be reduced to Eq.~\eqref{eq15:dir}, leading to the quantization of $\Gamma^{int}$. 

The above picture changes when anisotropy is turned on ($r \neq 0$, $\zeta = 0$), as the quantization of $\Gamma^{int}$ depends on the scale of $r$. This is because, unlike tilt, anisotropy affects eigenspinors, making the transition matrices depend on it. Since Chern number is defined as a surface integral about the source or sink of Berry flux, it is necessary to reduce the 3D integral in Eq.~\eqref{eqapp} into a surface integral, in order to connect it with the topological invariant. This manoeuvre is not possible for a finite $r$, and requires the limit $r \rightarrow 0$. We have shown the dependence of $\Gamma^{int}$ on $r$ in Fig.~\ref{fig:5}(c). As expected, quantization is lost if $r$ is increased. The fact that it follows a power-law decay in a log-linear plot, shows how sensitive the quantization is to the presence of anisotropy. This also hints at the fact that there must be a scale of $r$, beyond which quantization vanishes. To estimate this, we have computed $\Gamma^{int}$ for three different scales of $r$, as shown in Fig.~\ref{fig:5}(d). We notice that $\Gamma^{int}$ is in fact quantized, and equal to the Chern number, for $r \sim 10^{-3}$. Increasing $r$ by one order of magnitude by setting $r \sim 10^{-2}$, we find that although the behaviour of $\Gamma^{int}$ remains similar to the former case ($r \sim 10^{-3}$) in the lower frequency ranges, the response deviates dramatically with increasing $\omega$. Note that a small quantized plateau still forms at a lower frequency, despite the deviation. As we set $r=10^{-1}$, the response continues to fall off with a faster decay rate, and asymptotically tends to zero. This also explains the lack of quantization for $\alpha\neq0,2$ in the previous section, as the energy dispersion is anisotropic for those cases, which is analogous to the explicit addition of external anisotropies.

\begin{figure}[]
\centering
\includegraphics[width=\columnwidth]{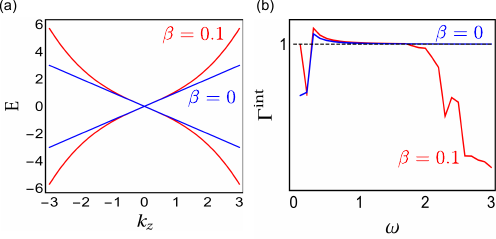}
\caption{\label{fig:6}
\textbf{(a)} Energy dispersion in the absence ($\beta=0$, blue line) and presence ($\beta=0.1$, red line) of the cubic term in Eq.~\eqref{eqcubicham}. The dispersion is linear in both cases at small energies. But much away from the node, the cubic term starts to dominate, and curves the dispersion. \textbf{(b)} Here we show $\Gamma^{int}$ (in the units of $ \frac{ \mathcal{E}^2}{2\, \pi \, \hbar^2}$) versus $\omega$, for $\beta=0$ and $\beta=0.1$. Contrary to the linear case, quantization for $\beta=1$ only exists at low frequencies, and eventually disappears as $\omega$ increases.}
\end{figure}

In addition to the linear terms in the components of $\mathbf{k}$, higher-order anisotropic terms may also appear within the energy resolution window. For example, the Hamiltonian~\cite{emil3}
\begin{align}
\label{eqcubicham}
	\mathcal{H}_{\text{cubic}} =  \hbar \, v_F \, \left[k_x \, \sigma_x + \, k_y \, \sigma_y + \, (k_z + \, \beta \, k_z^3) \, \sigma_z \right] \, , 
\end{align}
includes a cubic order correction with coefficient $\beta$. The dispersion of $\mathcal{H}_{\text{cubic}}$ is shown in Fig.~\ref{fig:6}(a) for $\beta= 0.1$. We find that the dispersion is approximately linear at low energies, but as the energy increases, the bands start to bend and deviate from the linear behaviour. The $\Gamma^{int}$ versus $\omega$ plot in Fig.~\ref{fig:6}(b) reveals that, in presence of the cubic term, the quantized plateau only exists within a frequency window, the range of which is set by the magnitude of $\beta$. This is because, Eq.~\eqref{eq:chern_appx} is no longer valid for higher frequencies, when the cubic term contribution becomes comparable to the linear terms.

\section{Summary and Outlook} 
\label{secend}

In this paper, we have described a probe (denoted by $\Gamma^{int}$) to detect the Chern number of the LBB in a 3D topological semimetal. This framework exploits the chiral nature of systems featuring Bloch bands with non-zero Chern numbers. However, unlike the DIR for probing Chern numbers in 2D systems \cite{goldman_dir}, the quantization of $\Gamma^{int}$ depends on the nature of the dispersions. For systems with linear dispersions, we found that $\Gamma^{int}$ is quantized above a critical frequency value $\omega_c$. It is important to note that this response is non-linear with respect to the amplitude of the circular drive, and it explicitly depends on interband transitions.

It is important to emphasize that the DIR response for 3D systems (which we denote by $\Gamma^{int}$), is different from the DIR formula for 2D systems (see Eq.~(8) of Ref.~\cite{goldman_dir}). First, $\Gamma^{int} (\omega)$ is a function of $\omega$ (i.e., integration over $\omega$ is not performed while deriving it). Second, $\Delta \gamma_{mn}^{\mu\nu} $ is computed for each 2D projection of the 3D system, lying in the $\mu\nu$-plane.
With $\lambda$ denoting the direction perpendicular to the 2D plane under consideration [cf. Eq.~\eqref{eq11}], we then multiply $\Delta \gamma_{mn} ^{\mu\nu}$ with the band velocity difference  $\partial_{k_\lambda } E_{mn}$, and integrate it over the entire 3D BZ to get the differential current along the $\lambda$ direction. This is followed by a summation over the three mutually perpendicular directions. This complicated expression happens to overlap with the expression of the Chern number of the LBB, at leading order, only for systems with isotropic dispersions, as we show in our derivations.

As one of the first examples, Bradlyn et al.~\cite{bradlyn} provided a comprehensive classification of space-group symmetry-protected band-crossings, and pointed out that multi-fold band touching points harness fermions with higher pseudospins. While pseudospin-1/2 fermions can occur at any generic points of the BZ, multiple-band-crossings (three, four, and eight) associated with higher pseudospins are only possible at high-symmetry points. In particular, both the pseudospin-1 and RSW fermions, considered in this paper, can only occur at high-symmetry points of the BZ. Additionally, the three-fold bandstructure of the pseudospin-1 fermions requires a nonsymmorphic structure of the space group, in contrast with the RSW fermions. As a part of the space-group classification, Bradlyn et al. proposed several candidate materials, supported by first-principle calculations, that can host higher-pseudospin fermions. For example, materials like $\mathrm{Pd_3Bi_2S_2}$ and $\mathrm{Ag_3Se_2Au}$ are predicted to have three-band-crossings near the Fermi level, while Fu et al.~\cite{hsieh} showed perovskite materials (with the chemical formula $\mathrm{A_3BX}$) can host pseudospin-3/2 fermions. Another set of proposals includes cubic silicides (ASi) and germanium materials (AGe), with A = Rh, Co. 

Recent ARPES measurements~\cite{takane} showed that CoSi host pseudospin-1 and double-Weyl fermions, at the centre and corner of the BZ, respectively. Similarly,
RhSi is reported~\cite{sanchez} to have both three-fold and four-fold band degeneracy at
the $ \Gamma$ and $R$ points. In such crystalline solids, circular dichroism can be probed by measuring orbital polarization through ARPES~\cite{schuler, unzelmann}.

The search for higher-pseudospin fermions have also expanded beyond conventional condensed matter crystals, and continues to explore non-electronic structures like cold-atomic and photonic frameworks. In particular, ultracold atoms trapped in optical lattices have received a lot of attention due to their ability to provide a highly tunable yet clean environment, making them one of the most promising platforms to realize topological semimetallic phases. Very recently, Wang et al.~\cite{wang21} realized a two-node ideal Weyl semimetal in ultracold atomic gases with synthetic spin-orbit coupling. This opens up the possibility of exploring various optoelectronic processes, especially DIR, which was originally proposed and implemented in 2D ultracold settings. 

Our proposal serves as a complementary platform, which can be used in conjunction with other probes \cite{Mandal_Weyl,ips_qbt_tunnel,ips3by2,ipsita-aritra,ipsfloquet,PhysRevB.103.195116,ips-herm} to map out various properties of semimetals. One future direction will be to investigate the behaviour of $\Gamma^{int}$ for other semimetals which have degenerate bands (e.g. Luttinger semimetals with quadratic band-touching points \cite{rahul-sid,ipsita-rahul,ips_qbt_tunnel,ipsfloquet,PhysRevB.103.195116,ips-herm}), or a mix of linear and non-linear dispersions depending on the momentum axis (e.g. multi-Weyl semimetals \cite{Mandal_Weyl,Mandal_2020,ipsita-aritra}). It will also be interesting to compute how disorder \cite{emil,emil2,rahul-sid,ipsita-rahul}, in the
presence of interactions \cite{roshtami,kozii,Mandal_2020}, affects the features of $\Gamma^{int}$.

\section*{Acknowledgments} 
We thank Kush Saha for useful discussions. The work is partially funded by the National Science Centre (Narodowe Centrum Nauki), Poland, under the scheme PRELUDIUM BIS-2 (grant number 2020/39/O/ST3/00973).

\begin{widetext}
\appendix

\section*{Appendix: Deriving the Transition Rate}

\
The Hamiltonian in Eq.~\eqref{eq:pert} can be expressed as
\begin{align}
\mathcal{H^{\prime}}_{\pm}(t)
& = 
\frac{\mathcal{E}} {\hbar \,\omega}
\left [ \left (e^{i\,\omega \,t} + e^{-i\,\omega\, t} \right) 
\frac{\partial \mathcal{H}_0}{\partial k_{\mu}}                  
\pm  \frac{e^{i\omega t} - e^{-i\omega t}}
{i}\, \frac{\partial \mathcal{H}_0}{\partial k_{\nu}} \right].
\end{align}
If we assume that the system is initally in the $n^{\text{th}}$ eigenstate, then the transition probability to the $m^{\text{th}}$ eigenstate at a finite time $t$ can be obtained using the time-dependent perturbation theory. For the harmonic perturbation in consideration in this paper, the transition rate to the leading order is given by
\begin{align}
 \gamma_{mn}^{\pm}(k_\mu,k_\nu,\omega)           
&  =\frac{2\, \pi}{\hbar} \,
	\left( \frac{\mathcal{E}}{\hbar \, \omega} \right)^2 \,                                                    
	\Bigg| \left\langle m \Bigg| \frac{\partial H_0}{\partial k_{\mu}} \pm \frac{1}{i} \, \frac{\partial H_0}{\partial k_{\nu}} \Bigg| n \right\rangle \Bigg|^2 
	\delta^t(E_{mn}- \hbar\, \omega) \,,
\quad \delta^t(\epsilon)   = 
\frac{
2\, \hbar  \, 
\sin^2 \left ( \frac{\epsilon\, t} {2 \, \hbar} \right)}
{  \pi\, t \,\epsilon^2}\,,	
\end{align} 
which is the Fermi's golden rule.
For sufficiently long timescales, the $\delta^t(\epsilon)$ can be approximated as $\delta(\epsilon)$, leading to Eq.~\eqref{eqgampm}.
Plugging all these in Eq.~\eqref{eq11}, and using the identity
\begin{align}
 |\mathcal{P}_{mn}^ \mu \pm i \, \mathcal{P}_{mn}^ \nu|^2 
 &= |\mathcal{P}_{mn}^ \mu|^2 + |\mathcal{P}_{mn}^ \nu|^2 
 \mp i \left (\mathcal{P}_{mn}^\mu \, \mathcal{P}_{nm}^\nu 
 - \mathcal{P}_{mn}^\nu \, \mathcal{P}_{nm}^\mu \right)  ,
\end{align}
we obtain
\begin{align}
\Delta \gamma_{mn}^{\mu \nu} (\mathbf k,\omega)
\equiv 
\frac{ \gamma_{mn}^{+}(k_\mu,k_\nu,\omega)  - \gamma_{mn}^{-}(k_\mu,k_\nu,\omega)    } 
{2}
=\frac{ 2\,\pi\,i} {\hbar}
	 \left( \frac{\mathcal{E}}{\hbar \, \omega} \right)^2                                                    
 \left (\mathcal{P}_{mn}^\mu \, \mathcal{P}_{nm}^\nu 
 - \mathcal{P}_{mn}^\nu \, \mathcal{P}_{nm}^\mu \right) 
\delta (E_{mn}- \hbar\, \omega)\,  ,
\end{align}
which leads to Eq.~\eqref{eq12}.

\end{widetext}

\bibliography{ref}
\end{document}